\documentclass{jpp}

\usepackage{graphicx}
\usepackage{natbib}

\ifCUPmtlplainloaded \else
  \checkfont{eurm10}
  \iffontfound
    \IfFileExists{upmath.sty}
      {\typeout{^^JFound AMS Euler Roman fonts on the system,
                   using the 'upmath' package.^^J}%
       \usepackage{upmath}}
      {\typeout{^^JFound AMS Euler Roman fonts on the system, but you
                   dont seem to have the}%
       \typeout{'upmath' package installed. JPP.cls can take advantage
                 of these fonts, if you use 'upmath' package.^^J}%
      }
  \else
  \fi
\fi

\ifCUPmtlplainloaded \else
  \checkfont{msam10}
  \iffontfound
    \IfFileExists{amssymb.sty}
      {\typeout{^^JFound AMS Symbol fonts on the system, using the
                'amssymb' package.^^J}%
       \usepackage{amssymb}%

      }{}
  \fi
\fi

\ifCUPmtlplainloaded \else
  \IfFileExists{amsbsy.sty}
    {\typeout{^^JFound the 'amsbsy' package on the system, using it.^^J}%
     \usepackage{amsbsy}}
    {}
\fi

\newsavebox{\astrutbox}
\sbox{\astrutbox}{\rule[-5pt]{0pt}{20pt}}

\title[Collisional behaviors of astrophysical collisionless plasmas]{Collisional behaviors of astrophysical collisionless plasmas}

\author[A. Bret]%
{A\ls N\ls T\ls O\ls I\ls N\ls E\ns B\ls R\ls E\ls T $^{1,2}$%
  \thanks{Email address for correspondence: antoineclaude.bret@uclm.es}
}

\affiliation{$^1$ETSI Industriales, Universidad de Castilla-La Mancha, 13071 Ciudad Real, Spain\\[\affilskip]
$^2$Instituto de Investigaciones Energ\'{e}ticas y Aplicaciones Industriales, Campus Universitario de Ciudad Real, 13071 Ciudad Real, Spain}

\date{?; revised ?; accepted ?. - To be entered by editorial office}
\begin{document}

\maketitle

\begin{abstract}
In collisional fluids, a number of key processes rely on the frequency of binary collisions. Collisions seem necessary to generate a shock wave when two fluids collide fast enough, to fulfill the Rankine-Hugoniot relations, to establish an equation of state or a Maxwellian distribution. Yet, these seemingly collisional features are  routinely either observed or assumed, in relation with collision\emph{less} astrophysical plasmas. This article will review our current answers to the following questions: How do colliding collisionless plasmas end-up generating a shock as if they were fluids? To which extent are the Rankine-Hugoniot relations fulfilled in this case? Do collisionless shocks propagate like fluid ones? Can we use an equation of state to describe collisionless plasmas, like MHD codes for astrophysics do? Why are Maxwellian distributions ubiquitous in Particle-In-Cell simulations of collisionless shocks? Time and length scales defining the border between the collisional and the collisionless behavior will be given when relevant. In general, when the time and length scales involved in the collisionless processes responsible for the fluid-like behavior may be neglected, the system may be treated like a fluid.
\end{abstract}

%\begin{PACS}
%Authors should not enter PACS codes directly on the manuscript, as these must be chosen during the online submission process and will then be added during the typesetting process (see http://www.aip.org/pacs/ for the full list of PACS codes)
%\end{PACS}

\section{Introduction}
Astrophysical plasmas are usually collisionless. It means that the mean free path for binary Coulomb collisions is usually much larger than the typical size of the system. Another way of putting it consists is stating that the average time between electronic Coulomb collisions $\propto N_D\omega_p^{-1}$, where $N_D$ is the number of electrons per Debye sphere \citep{spitzer}, is much larger than the other time scales of the problem. In collisionless plasmas with $N_D\rightarrow\infty$, the Spitzer collision frequency goes to zero, and many phenomena can be studied by completely neglecting it.

Another important consequence of collisionless-ness is that collisions can no longer thermalize the system on short time scales. Because binary collisions are responsible for the convergence of the distribution function toward a Maxwellian \citep{Chandrasekhar} on a ``Spitzer time scale'', the absence of collisions suppresses the most common thermalization agent. It may therefore be problematic to use an equation of state for the medium, since the very existence of such an equation assumes at least a local thermal equilibrium. Yet, astrophysicists make an intensive use of MHD codes requiring an equation of state \citep{Hawley1984,Zeus1992,OlecMHD2014}. To which extent is such a strategy appropriate? Although we don't have any definite answer so far, progresses have been made in recent years pointing toward faster-than-Spitzer thermalization mechanisms, which could justify treating collisionless plasmas like collisional fluids in appropriate settings.

Collisionless shocks illustrate also the fluid-like behavior of collisionless plasmas. When two collisional fluids collide fast enough, or when a large amplitude sound wave is launched in one fluid, a shock wave is generated \citep{Zeldovich}. At the fluid, macroscopic, level, this wave represents a traveling discontinuity. And macroscopic quantities like density, temperature or pressure on both sides of the discontinuity obey the so-called Rankine-Hugoniot (RH) relations. It turns out that the encounter of two collisionless plasmas also results in the formation of a shock, although both plasmas never ``hit'' each other, but start passing through each other instead. Furthermore, simulations show a fast thermalization (or at least a thermal component) of the medium when crossing the front \citep{Spitkovsky2008a,lapenta2007,lapenta2009,Sironi2011}. The RH conditions are satisfied to a reasonable accuracy, whether for simulations or for the earth bow shock \citep{Farris1992}.

We thus find several instances of fluid-like behavior of collisionless plasmas. Whether it be the issue of thermalization or the shock problem, it seems that some collisionless mechanisms can achieve what binary collisions cannot. We will here review the current status of our knowledge in these matters.

\section{Fluid-like shock physics}
When a shock propagates, upstream particles need to dissipate part of their kinetic energy for the fluid to slow down in the downstream. For a collisional fluid, this is achieved through an increase of the number of inter-particle collisions. This is why the shock front is a few mean-free-paths thick \citep{Zeldovich}, that is, a discontinuity at the fluid level. Let us now turn to the earth bow shock. Here, we know from in situ measurements that its front is about 100 km thick \citep{PRLBow1,PRLBow2}. But the proton mean-free-path at the same location is comparable to the sun-earth distance \citep{kasper2008}. Or turn to the supernova remnant SN 1006. Here, the shock front is about 0.04 pc ($1.2\times 10^{17}$ cm) while the mean-free-path is 13 pc ($4\times 10^{19}$ cm) \citep{bamba2003,takabe2008}.

It is thus clear that the mechanism ensuring dissipation at the shock front cannot come from binary collisions for these astrophysical shocks. Indeed, there was a time when the mere existence of such shocks was simply questioned \citep{Sagdeev1966,Sagdeev1991}. We shall now see how they are formed, and how they result in structures fulfilling the RH jump conditions.

Although this review is not about collisionless shocks per see (see \cite{Treumann2009,Treumann2011,balogh2013physics} for such reviews), we briefly mention in Appendix \ref{sec:zoo} the various kind of collisionless shocks one can encounter.

It is worth specifying that the fluid-like behaviors we're commenting relate only to unmagnetized shocks in pair and electron/ion plasmas. Yet, to our knowledge, no simulation of colliding plasmas that would produce a shock for collisional fluids, has failed so far to produce one with collisionless plasmas \citep{SironiPC}.

We now turn to our first example of fluid-like behavior. How does the encounter of two unmagnetized, collisionless plasmas, result in the formation of a shock, like it happens for fluids? The detailed theory of collisionless shock formation we are about to review for two simple cases will help understand.

\subsection{Unmagnetized pair plasmas}\label{sec:shockpair}
The setup consisting of two identical colliding  pair plasmas is attractive for several reasons:
\begin{enumerate}
  \item The absence of particles of different mass renders both theory and simulation much simpler.
  \item By virtue of this mass issue, pair plasmas don't exhibit a Debye sheath at their border \citep{gurnett2005}. In case two of them collide, they will always have enough energy to start interpenetrating.
  \item Pair plasmas are directly relevant to some astrophysical settings \citep{usov1992,Broderick2012,Sironi2014}, and available in the laboratory \citep{Chen2010,Sarri2013}. In addition, experiments mimicking pair plasmas can be performed with positively and negatively charged $C_{60}$ molecules \citep{PairC601,PairC602}.
\end{enumerate}

\begin{figure}
\begin{center}
\includegraphics[width=0.9\textwidth]{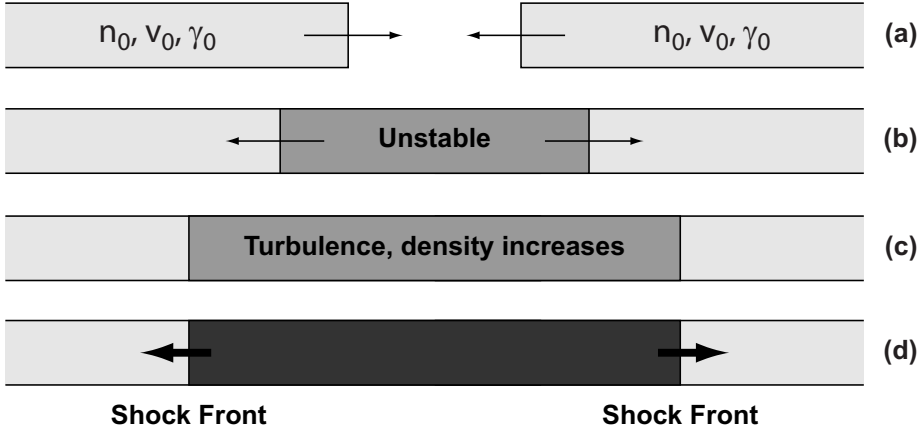}
\end{center}
\caption{Formation of a collisionless shock via the encounter of 2 pair plasmas. (a) The 2 plasmas shells are heading toward each other. (b) They start interpenetrating, and the overlapping region turns unstable. (c) Instability saturates, generating a strong enough turbulence to block the incoming flow. The density in the central region increases until RH conditions are fulfilled. (d) Two shocks form and propagate. For simplicity, identical plasmas are shown.} \label{fig:1}
\end{figure}

In the absence of binary collisions, two identical plasmas heading toward each other at velocity $\pm \mathbf{v}_0$ don't collide (see figure \ref{fig:1}). Instead, they simply pass through each other. How is it then that a shock eventually forms? The overlapping region could grow indefinitely, at least until the Spitzer time, if it were not for counter-streaming instabilities \citep{Kahn1958,Buneman1964}. Being formed by two counter-streaming plasmas, instabilities will grow in this region, saturate, and generate a turbulence. Though many instabilities do grow, as explained in Appendix \ref{sec:weibel}, the Weibel instability is the fastest growing one in these circumstances, which is why these shocks are called ``Weibel mediated''.

 If the randomization length of the incoming flow $L_R$, is smaller than the size of the central region at saturation $L_S$, the plasma which keeps entering the overlapping region stops there \citep{BretPoP2013,Bret2014}. As a result, the bulk velocity in the central region is now zero.

At this stage, the system presents therefore a macroscopic discontinuity in velocity space, but the central density is still the sum of the densities of the two shells. The time needed to get to this stage from the first contact of the plasmas is given by,
\begin{equation}\label{eq:t:sat}
  \tau_s=\delta_m^{-1}\Pi,
\end{equation}
where $\delta_m$ is the growth rate of the dominant instability (see Appendix \ref{sec:weibel}) and $\Pi$ the number of exponentiations until saturation. Provided the soon-to-form shock will have a density jump $\Gamma$ given by the RH relations (see section \ref{sec:rh}), the density jump will now increase from 2 to $\Gamma$. If it took a time $\tau_s$ to bring enough material in the central region for a density jump 2 to build-up, the time needed to reach $\Gamma$, namely the shock formation time, will be
\begin{equation}\label{eq:t:form}
  \tau_f \sim \tau_s + (\Gamma-2)\tau_s = (\Gamma-1)\tau_s.
\end{equation}

Therefore, provided one is interested in time scales longer than $\tau_f$ and length scales larger than $L_S\sim 2 v_0\tau_s$, we recover a fluid-like behavior: the two shells ``collide'', and two opposite shocks are generated in each one.

\subsection{Unmagnetized  electron/ion plasmas}\label{sec:shockei}
Things are different when particles of various masses are involved, especially by virtue of the large proton-to-electron mass ratio. On the one hand, various instabilities may develop on their proper time scale, so that the building up of a turbulent region is more involved than for pair plasmas. On the other hand, the Debye sheaths  bordering these plasmas can interfere at low energies, and alter the shock formation.

For low impacting energies, i.e., if the initial kinetic energy of our identical shells is not high enough for protons to overcome the potential barrier of the other shell, a collisionless electrostatic shock is formed within a time \citep{Forslund1970,Dieckmann2013,Stockem2014},
\begin{equation}\label{eq:t:form2}
  \tau_f \sim 10\gamma_0^{3/2}\omega_{pi}^{-1}.
\end{equation}
Note that the impacting velocity may be relativistic for high-enough temperatures, hence the Lorentz factor $\gamma_0$ in the equation above. Also, the dominant time-scale is now the protonic one, which is why the formation time scales like the proton plasma frequency $\omega_{pi}$. Protons also govern the width of the shock front, which is eventually a few $c/\omega_{pi}$ thick.

As the impacting energy increases, a transition region has to be crossed which is not straightforward to describe. Here, we can refer the reader to \cite{Stockem2014}. Then, for high enough energies, roughly $(\gamma_0-1)m_ec^2\gg k_BT_e$, we recover a situation similar to the pair plasmas case. Because there is enough energy to overcome the potential barriers, the two plasma shells start to overlap. The counter-streaming electrons first turn Weibel unstable, and their dominant instability saturates. The counter-streaming protons then turn Weibel unstable, and the relevant system at this stage consists in two opposite proton streams over a bath of hot electrons \citep{Shaisultanov2012,Davis2013}.

Denoting again $\delta_m$ the growth rate of the dominant instability of this system, we can repeat the reasoning done previously for pair plasmas, and derive an equation similar to (\ref{eq:t:form}) for the formation time. Note however that some theoretical work is still needed to fully explain the simulations. One reason is that the growth rate $\delta_m$  depends on the temperature of the hot electrons bath. It appears from the simulations that they are heated to a temperature much higher than their initial streaming energy \citep{Gedalin2012,Plotnikov2013,Davis2013}. Although various theories have been proposed to explain this anomalous electron heating, the situation is not clear yet.

To conclude this section on shock formation, we can say that a collisionless shock forms when two collisionless plasmas collide, over time and length scales given by the equations above. We therefore recover a fluid-like behavior, though the underlying physics completely differs from the collisional case.

\begin{figure}
\begin{center}
\includegraphics[width=0.9\textwidth]{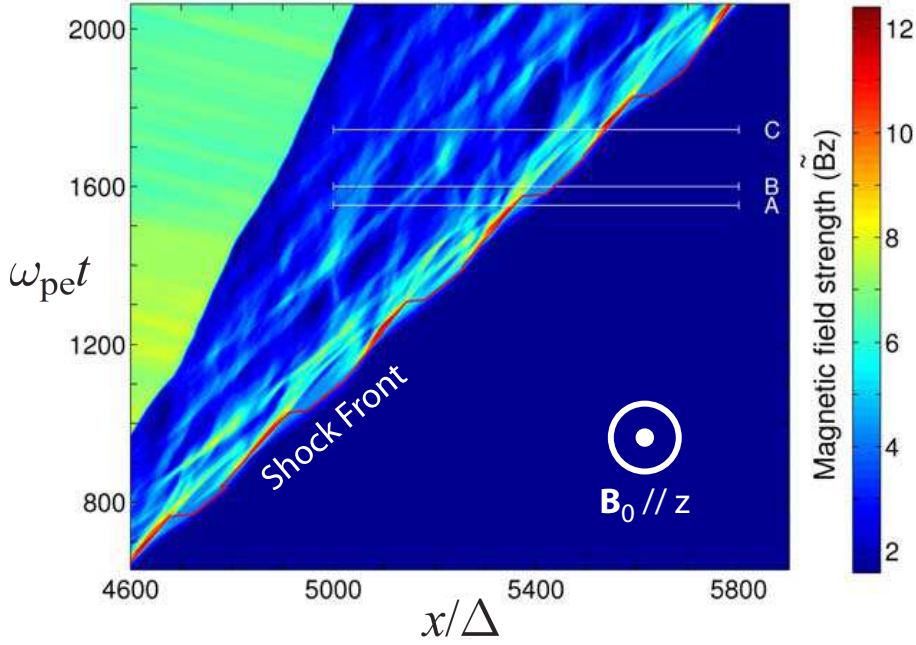}
\end{center}
\caption{Non-stationarity of a collisionless shock. The front propagates from left to right. $\Delta$ is the numerical grid size. The evolution of the field at the front evidences the non-stationarity of the shock. The upper-left green rectangle pertains to the moving piston generating the shock. From \cite{Lembege2009}.} \label{fig2}
\end{figure}

\subsection{Shock stability and stationarity}\label{sec:shocksatb}
Once the shock has formed, the RH conditions determine its velocity, both in the fluid and the collisionless case (see section \ref{sec:rh}). A key issue which has been established long ago in the fluid case is the one of the stability of the shock front. Whether we deal with the non-relativistic \citep{Dyakov1954,Kontorovich1958,Wouchuk2004} or the relativistic case \citep{Anile1987}, a corrugation applied to the shock front in an ideal fluid does not grow. Though instabilities have been evidenced for shocks in substances described by a van der Waals equation of state \citep{Bates2000}, shocks in ideal fluids are stable.

What about the collisionless case? To our knowledge, no theoretical work has proved so far the front stability or instability against all kinds of perturbations. The challenge here is that electromagnetic perturbations have to be considered, in addition to the shock front corrugation. Regarding stability against electromagnetic fluctuations for electrostatic shocks, \cite{Stockem2014} could relate it to the electrostatic/electromagnetic shock transition.

The process of collisionless shock re-formation could indicate a departure from the fluid stability, or at least stationarity. This phenomenon was first noticed in 1D simulations \citep{lembege1987}, before it was observed in 2D \citep{Lembege1992} and in nature on the earth bow shock \citep{Mazelle2010}. Also referred to as shock ``non-stationarity'' \citep{Lembege2004}, this process occurs for collisionless shocks propagating quasi-perpendicularly to an ambient magnetic field.

This self-reformation is due to reflected ions which accumulate in front of the shock. It results in a non-stationary evolution of the shock front on a time scale given by the upstream ions gyrofrequency $\sim\omega_{ci}^{-1}$,  and a length scale about the ions gyroradius $\lambda_i$, as evidenced on figure \ref{fig2}. For this figure, the shock has been generated by a piston moving rightward, and visible on the plot through the upper-left green rectangle. Like fluid shocks \citep{landau6}, collisionless shocks can also be generated by a moving piston.

Non-stationarity is not the fruit of some external factor. It is part of the proper shock dynamics. Provided one is interested in long enough time scales ($\gg \omega_{ci}^{-1}$), and large enough length scales ($\gg \lambda_i$), a non-stationary shock behaves like a fluid one. So far, it has been found stationary.

\subsection{Downstream ion-electron thermal equilibration}\label{sec:equi_shock}
The issue of  ion-electron thermal equilibration in the downstream of some collisionless shocks could equally imply some departure from the collisional behavior\footnote{See also \citep{Raymond2010}.}. Based on observations of some Supernova remnants, galaxy cluster shocks and the terrestrial and Saturnian bow shocks, \cite{Ghavamian2013} found that the ratio $T_e/T_i$ of the electron to the ion temperatures falls like $\mathcal{M}^{-2}$, where $\mathcal{M}$ is the Mach number. An analytical two fluids model established $T_e/T_i\sim 1$ for $\mathcal{M}\lesssim 2$, $T_e/T_i\sim m_e/m_i$ for $\mathcal{M}\gtrsim 60$, and $T_e/T_i\propto \mathcal{M}^{-2}$ in the intermediate regime \citep{Vink2014}. Little is known however on the microphysics involved in equilibrating the ions and electrons temperatures \citep{Matsukiyo2010}. Time will therefore be needed to draw definite conclusions on the collisional status of this feature. At any rate, the ultimate collisionless picture could be equivalent to a \emph{two}-fluids model, instead of a single-fluid one.

\section{Rankine-Hugoniot relations and equation of state}\label{sec:rh}
The foundations of the RH jump conditions reside in the conservation of matter, momentum and energy. For the fluid case, they are simply derived writing that these quantities are conserved when crossing the shock front \citep{Zeldovich}. To which extent can they remain valid for a collisionless shock?

\begin{figure}
\begin{center}
\includegraphics[width=0.9\textwidth]{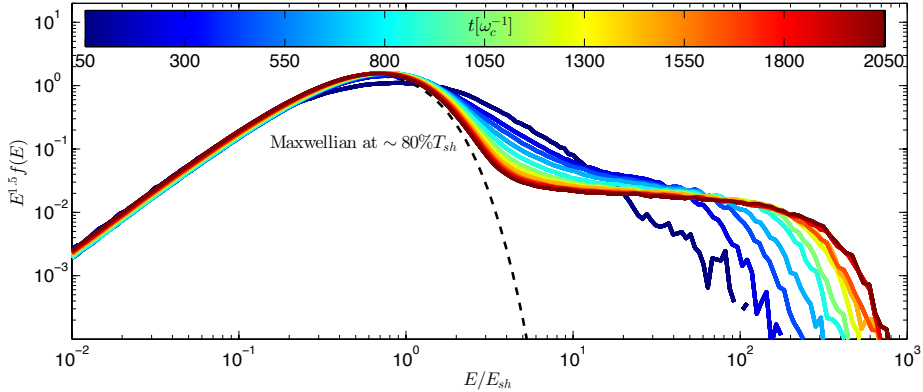}
\end{center}
\caption{Downstream ion energy distribution in terms of time in a shock simulation. A Maxwellian is clearly evidenced, together with a power-law tail extending to larger energies at later times. The spectrum is multiplied by $E^{1.5}$, to emphasize the agreement with the scaling predicted by the acceleration mechanism. Time is measured in unit of $\omega_c^{-1}$, the ion cyclotron frequency in the external magnetic field. The energy $E$ is normalized to $E_{sh}=\frac{1}{2}mv_{sh}^2$, where $v_{sh}$ is the upstream fluid velocity in the downstream reference frame. From \cite{Caprioli2014}.} \label{figmax}
\end{figure}

Since matter carries all conserved quantities (we exclude radiative shocks), we need to make sure all the matter upstream ends up downstream in the collisionless case. This is verified, but not perfectly because some particles from the upstream can be reflected by the shock front, and come back upstream. Also, other particles may be accelerated near the front, escaping the system nearly isotropically. A numerical investigation of these effects demonstrated a departure from RH of a few percent (see \cite{Stockem2012} and concluding remarks). For shocks which are efficient particle accelerators\footnote{In the magnetized regime, it seems these are the quasi-parallel shocks, where the angle $\theta_B$ between the shock front normal and the external field is not ``too close'' to $\pi/2$ \citep{Lemoine2006,Caprioli2014NuPh}. For non-relativistic shocks, \cite{Caprioli2014NuPh} gives $\theta_B\lesssim \pi/4$.}, up to  10\%-20\% of the upstream kinetic energy can be converted to energetic particles \citep{Caprioli2014}. As a result, the downstream is found 20\% cooler than it would be for a fluid (see figure \ref{figmax}).

We can therefore state that up to a $\sim 10-20\%$ accuracy, all the matter upstream of a collisionless shock passes downstream, together with the momentum and energy it carries. Yet, the RH conditions under their usual form need more than the conservation laws. They need an equation of state of the form,
\begin{equation}\label{eq:eos}
P=(\gamma-1)U,
\end{equation}
where $P$ is the pressure, $\gamma$ is the adiabatic index and $U$ the energy density of the medium. As showed in Appendix \ref{appA}, such an equation of state does not require a Maxwellian distribution. An isotropic distribution function, in the reference frame where the plasma is a rest, is enough\footnote{I thank Guy Pelletier for pointing this out to me.}. In simulations, the upstream distribution is an input usually considered isotropic. Regarding the downstream distribution, turbulent magnetic fields are present both in the downstream of electromagnetic and electrostatic shocks \citep{StockemPRL2014}, that are capable of nearly isotropizing the distributions. It is therefore not surprising to observe the near fulfillment of the RH conditions in simulations of collisionless shocks (see for example \cite{Spitkovsky2008a,SironiApj}).

Yet, it seems we not only observe isotropic distributions. We observe Maxwellian distributions. Some processes must therefore be at work to operate such a collisionless thermalization. This is the topic of the next section.

\section{Collisionless thermalization}\label{sec:therm}
Figure \ref{figmax} displays the time evolution of the downstream ion energy distribution in a shock simulation. A remarkably stable Maxwellian component is clearly visible at low energies. At $t=2500\omega_c^{-1}$, its temperature is 80\% of what it would be for a fluid shock. Another component, a power-law with a cut-off, is  present as a result of particle acceleration \citep{Fermi1949,Krymskii1977,Blandford78,Bell1978a}.

Where does this Maxwellian come from? As already mentioned, the magnetic fields which are present either in the shock front or in the near downstream, can explain isotropization, but thermalization, i.e. Maxwellianization, is achieved over time scales far shorter than the Spitzer one. In the absence of collisions, the distribution function is solution of the Vlasov equation,
\begin{equation}\label{eq:vlasov}
 \frac{\partial f}{\partial t} +
 \mathbf{v}\cdot\frac{\partial f}{\partial\mathbf{r}} +
 q\left(\mathbf{E}+\frac{1}{c}\mathbf{v}\times\mathbf{B}\right)\cdot\frac{\partial f}{\partial\mathbf{p}} = 0,
\end{equation}
which does not demand a Maxwellian. What is then its origin?

It is remarkable to notice that the same enigma equally arises when studying the distribution of stars' velocity in a galaxy. As noted by \cite{Lynden1967MNRAS}, ``The remarkable regularity in the light distribution in elliptical galaxies suggests that they have reached some form of natural equilibrium. However, estimates of the normal star-star relaxation show that it is too weak to establish equilibrium in the time available''. The concept of ``violent relaxation'' introduced by Lynden-Bell to explain this phenomena, is still debated today \citep{Lynden2005MNRAS,binney2011}.

Coming back to plasma physics, the anomalous occurrence of Maxwellian distribution functions has been known for nearly one century as ``Langmuir's Paradox''. In 1925, Irving Langmuir observed the persistence of a Maxwellian distribution in an experiment where ``the number of collisions of the electrons with each other and with atoms were far too few to maintain the observed distribution''  \citep{Langmuir1925}. The solution to this problem would require decades of efforts, and \cite{gabor1955} even referred to it as ``one of the worst discrepancies known in science''\footnote{See \cite{baalrud2010kinetic} for a detailed history of the Langmuir's paradox.}.

\cite{Kadomtsev1965} discussed this process in terms of turbulent plasma heating. This author invoked the strong turbulence generated by beam plasma instabilities to explain collisionless thermalization. Yet, no definite theory could be laid out at that time, as ``the real picture of the excitation of waves by powerful beam and the interaction of the waves with the particles and with one another may turn out to be much more complete, and to require a considerable development of the theory for its complete description'' (\cite{Kadomtsev1965}, p. 131).

Without developing a detailed theory, \cite{hoyaux1968arc} noted that the randomizing effect of plasma turbulence could be modeled considering particles are subject to a macroscopic force and a microscopic, stochastic, one. The resulting Langevin equation models a Brownian motion in velocity space which provides a Maxwellian distribution, provided the stochastic force has a Gaussian distribution \citep{Chandrasekhar,Dunkel2006}. Such is the way \cite{Dieckmann2006} give account of the Maxwellians found in the beam-plasma simulation they study.

After more than 80 years of research, a breakthrough has recently been made by \cite{Baalrud2009,Baalrud2010}. Starting either from the BBGKY hierarchy \citep{cercignani1976,Ichimaru} or from the Klimontovich equation \citep{klimon}, these authors derived an ``instability-enhanced'' collision term for Eq. (\ref{eq:vlasov}) which accounts for wave-particle scattering due to plasma instabilities. This non-relativistic theory neglects the magnetic field produced by the charges in motion. When applied to the conditions of Langmuir's experiment, the enhancement of electron-electron collisions produced by the ion-acoustic instabilities produces a convergence toward ``nearly a Maxwellian'' (more on this in the conclusion) 100 times faster than the Spitzer time, explaining the paradox. The number 100 depends on various factors, like the plasma density and temperature, the experimental setup and the kind of unstable modes excited.

Noteworthily, convergence to a Maxwellian requires a collision operator with both a drag-term and a diffusion-term. The quasi-linear theory like the one developed by \cite{Vedenov1963} only displays a diffusion term. It misses a drag term to  describe adequately the convergence to a stationary distribution. Regarding the Fokker-Planck equation, it is known it reduces to the quasi-linear theory when the two terms of its collision operator are computed neglecting the recoil of the particles' scatterers \citep{Blandford1987,Peeters2008}. As such, it cannot describe either the convergence toward a stationary distribution.

At any rate, an instability generated turbulence seems required to achieve fast thermalization. This is related to the fact that the instability-enhanced collision operator fulfills the $\mathcal{H}$-theorem, and vanishes for a definite distribution. Let us remind that in contrast, a stable Vlasov plasma, when perturbed even non-linearly, evolves at constant entropy \citep{landau10} and comes back to a different distribution function than the one it had in the first place \citep{Villani2014}. Simply put, Landau damping does not restore a Maxwellian\footnote{Which is the reason behind plasma echo (\cite{Gould1967}, \cite{landau10}, p.141).}.

Turbulent settings, like the ones arising from the development of plasma instabilities, favor the collisional behavior of collisionless plasmas. One the one hand, turbulence leads to an isotropic medium. On the other hand, plasma turbulence is likely to produce the kind of enhanced collisionality described by  \cite{Baalrud2009,Baalrud2010}, so far in the non-relativistic, electrostatic case.

\section{Instabilities and related phenomena}\label{sec:insta}
We now comment on parallels between some instabilities and related phenomena in the collisional and collisionless regime. Clearly, in order to compare any collisionless behavior to some fluid counterpart, the instability under scrutiny must exist in the fluid world. It means it must exist in a system which can be modelled by a single fluid at a given position in space. Excluded then are the plasma streaming instabilities, like two-stream or filamentation, which by definition involve the counter-motion of several fluids at the same place. Note than these can be described quite successfully by a \emph{two}-fluids model \citep{BretPoPFluide}. Yet, we here restrict to the parallel with a single fluid.

In the case of the fluid \emph{Kelvin-Helmholtz instability}, a corrugation at the interface between two shear flows grows as a result of vorticity dynamics on each side of the disturbance \citep{batchelor2000}. For flows of the same density with velocities $\pm \mathbf{V}$, the fluid growth rate for disturbances with wave vector $\mathbf{k}\parallel \pm \mathbf{V}$ is $\delta_{KF, F}=\frac{1}{2}kV$ \citep{landau6}. The same instability has been studied both theoretically and numerically for two semi-infinite collisionless plasmas drifting in opposite directions \citep{Gruzinov2008,Grismayer2013,Alves2014}. For $k\ll \omega_p/V$, the growth rate is simply $2\delta_{KF, F}=kV$. But here, the underlying mechanism is similar to the two-stream instability, which has also $\delta=kV$ for large wavelengths \citep{Mikhailovskii1}. The coincidence of the growth rates is striking, in spite of apparently different instability mechanisms. It is probable that some deeper physical connection can be found between the fluid and the collisionless plasma KH instabilities.

Instabilities have been also considered in connection to the \emph{collisional behavior of ions in incompressible, collisionless turbulence}. Again within shear flows, here magnetized, a turbulence is generated out of the growth of the firehose and mirror instabilities \citep{Kunz2014}, or out of the growth of the mirror and the ion cyclotron instabilities \citep{Riquelme2014}. A collisional behavior is retrieved. For example, whether one deals with the firehose or the mirror instability, the instability itself generates an effective scattering rate maintaining the pressure anisotropy to the level of a collisional plasma  \citep{Kunz2014}.

The \emph{magnetorotational instability} is of great relevance for the physics of accretion disks. It has been studied first within the MHD framework \citep{Balbus1998}. PIC simulations evidenced a behavior very similar to MHD simulations \citep{Riquelme2012}. Theoretical works found identical kinetic and MHD stability criteria, while the growth-rates agreement depends on the plasma magnetization \citep{Quataert2002}. Indeed, the MHD-like behavior of the MRI is tightly related to the presence of kinetic plasma instabilities that suppress the growth of pressure anisotropies, thus mimicking the effect of collisions \citep{Sharma2006}.

Also relevant to the physics of accretion disks is the problem of \emph{ion-electron temperature equilibration}. In low-luminosity accretion disks, the ions are hotter than the electrons \citep{Yuan2014}. In such collisionless environments, binary collisions could still equilibrate the temperatures, but on the extra-long Spitzer time scale. Hence, \cite{Sironi2014a,Sironi2014b} investigated how ion driven plasmas instabilities can transfer energy to the electrons. In this respect, it seems there is ``solid evidence'' that the ion cyclotron instability tend to equilibrate ionic and electronic temperatures, if the latter are initially much cooler than the former ($T_{0e} \lesssim 0.2 T_{0i}$). Further works will be needed to assess to full parameter range, allowing for example to introduce an extra electron heating term in MHD codes.

\section{Conclusion}
Collisionless plasmas behave quite like collisional fluids in some conditions. Like in fluids, shocks can be generated launching them against each other. To the extent that the downstream distribution function is reasonably isotropic, an equation of state can be used, and RH jump conditions applied with good accuracy. As it propagates, the shock may reform and/or accelerate particles, so that its stationarity is only achieved up to certain time and length scales.

The progressive building-up of the high energy power-law tail observed on Fig. \ref{figmax} eventually reaches saturation \citep{Caprioli2014ApJL}. For weakly magnetized shocks with $\sigma < 10^{-1}$ (see Eq. \ref{eq:sigma}), \cite{Sironi2013} found that the maximum energy of the accelerated particles grows like $t^{1/2}$, before it saturates when these particles' energy reaches about 10\% of the flow. The 20\% departure from the downstream RH temperature could thus be an upper-bound. Indeed, some Supernova Remnants have been around for millennia\footnote{SN 1006, already mentioned, was born in AD 1006 \citep{murdin1985}. The Supernova Remnant GSH 138–01–94 is 4.3 million years old \citep{Stil2001}.}, suggesting they could have reached a kind of steady state. Even if \cite{Sironi2013} and  \cite{Caprioli2014ApJL} show how this can happen for the shock they study, the same conclusion is yet to reached for all kinds of collisionless shocks.

Also, the ``faster-than-Spitzer'' Maxwellianization observed in experiments or simulations has been explained, so far for the non-relativistic electrostatic case only, as an instability-enhanced collisional effect. The system is still collisionless, but particles scattering on an instability generated turbulence, drive a fast thermalization.

In spite of these numerous fluid-like features, some words of caution seem appropriate:

\begin{itemize}
  \item Using an equation of state requires an isotropic distribution (see Appendix \ref{appA}). Yet, such isotropy can be jeopardized in strongly magnetized settings. In this respect, \cite{Chew1956} tried to derive the MHD equations, starting from the Vlasov equation with a large Lorentz force. Their attempt was dimmed ``not entirely successful'' due to the anisotropy brought by the magnetic field lines.
  \item Even for the instability-enhanced thermalization, the distribution function converges only ``nearly to a Maxwellian'' \citep{Baalrud2010}, because the collision operator vanishes for distributions only close to a Maxwellian. This could be the reason why an electrostatic shock produces a flat-top electron distribution in 1D. In 1D-electrostatic the distribution is constrained by the fact that phase space paths cannot intersect. A 2D simulation with the same initial conditions generates ion acoustic turbulence downstream of the shock and that one produces rapidly an electron distribution with a single maximum that looks Maxwellian-like \citep{dieckmann2013pa,Dieckmann2013}.
\end{itemize}

Instability-enhanced collision frequency seems thus the key to fast thermalization. Now that Langmuir's paradox has been solved, one can wish further works will characterize the turbulence needed to generate Maxwellian distributions, beyond the initial Langmuir's experimental setup. Since (quasi) thermalization is observed in PIC simulations beyond the scope of \cite{Baalrud2009,Baalrud2010}, it could be possible to extend the proof to the relativistic regime, or to the case of a full electromagnetic turbulence.

Beyond this, the similarities between the fluid and the collisionless Kelvin-Helmholtz instabilities allow to think that some general physical principles could be at work behind these apparently different processes. On the long term, maybe such principles will allow to draw a clear quantitative line between the collisionless and the fluid behaviors of diluted plasmas.

\section{Acknowledgements}
The author acknowledges the support from grant ENE2013-45661-C2-1-P from the Ministerio de Educaci\'{o}n y Ciencia, Spain. Thanks are due to Didier B\'{e}nisti, Eric Blackman, Damiano Caprioli, Mark Dieckmann, Laurent Gremillet, Thomas Grismayer, Bertrand Lemb\`{e}ge, Ramesh Narayan, Guy Pelletier, Aleksander Sadowski and Gustavo Wouchuk for inputs and enriching discussions.

\appendix

\section{Shock zoology}\label{sec:zoo}
\begin{figure}
\begin{center}
\includegraphics[width=0.9\textwidth]{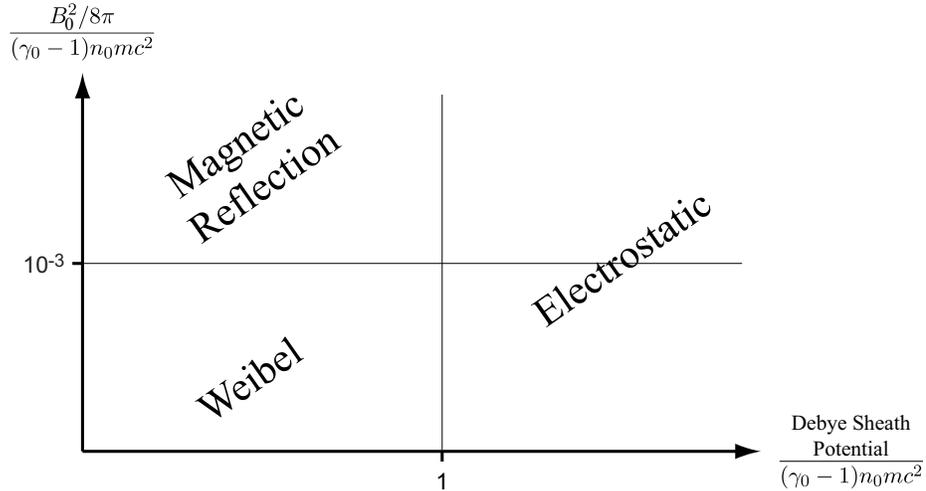}
\end{center}
\caption{Zoology of collisionless shocks. Different kind of shocks develop according to the magnetization, and the height of the Debye sheaths bordering the colliding plasmas. This classification could be enriched considering even more ingredients (see text).} \label{fig:zoo}
\end{figure}

Because plasmas are composed of charged particles, they can interact with electromagnetic fields. This feature, combined with the Debye sheath likely to border a plasma, introduces a rich variety of cases when considering the encounter of two plasmas. The typology related to shocks generated by colliding 2 plasmas is schematically represented on figure \ref{fig:zoo}.

\begin{itemize}
  \item The first important parameter is the magnetization. At low magnetization \citep{Niemiec2012}, namely
  \begin{equation}\label{eq:sigma}
  \sigma \equiv \frac{B_0^2/8\pi}{(\gamma_0-1)n_0mc^2} \lesssim 10^{-3},
  \end{equation}
  randomization of the upstream at the front is provided by the magnetic field generated by the Weibel instability (see section \ref{sec:weibel}). For high magnetization, it rather comes from reflection on the compressed field at the front. In this respect, besides the field compression due to the RH conditions including the magnetic energy \citep{tidman1971}, some plasma instabilities can provide an additional enhancement of the field at the front \citep{Bell2004}.
  \item Then comes the height of the Debye sheath, compared to the kinetic energy. Because ions are much heavier than electrons, the electronic density does not drop to zero like the ionic one at the border of an electron/ion plasmas. Such plasmas are instead bordered by a Debye sheath displaying a potential jump $\sim k_BT_e/q$ high, where $T_e$ in the electronic temperature and $q$ the elementary charge \citep{gurnett2005}. If the initial kinetic energy of the colliding plasmas is high enough, that is, $(\gamma_0-1)n_0mc^2 \gg k_BT_e$, both interpenetrate and an electromagnetic shock will be generated. On the contrary, if the interaction is mediated by the one of the Debye sheaths, an electrostatic shock is formed. Although not formally proven, this is probably the reason why electrostatic shocks cannot exist in pair plasmas with equal temperatures for electrons and positrons \citep{Oohara2003,Verheest2005}. But if electrons and positrons have different temperatures, a Debye sheath can form, which is probably connected to the possibility of electrostatic shocks in such plasmas \citep{Dubinov2006}\footnote{Strictly speaking, these references deal with solitary waves, not shock waves. Yet, as nicely explained by Sagdeev, both structures are closely related since a soliton is a large amplitude wave leaving the medium unchanged after its passage, while a shock does affect the medium. Simply put, ``soliton + dissipation = shock'' \citep{Sagdeev1966,Krall1997}.}.
\end{itemize}

Although their formation mechanism slightly differ (see sections \ref{sec:shockpair} \& \ref{sec:shockei}), shocks in pair or electron/ion plasmas can therefore be classified according to the same typology when focusing on the Debye sheath potential. Still, figure \ref{fig:zoo} falls short of exhausting every possibility. To do so, we would have to evoke the encounter of two different plasmas (pair vs. electron/ion, different densities\ldots \cite{Dieckmann2014}). We could also contemplate plasma shells including more than 2 species, like electron + positron \& proton, or electron + ion \& proton \citep{DieckmannApJ2009}.

For magnetized shocks, yet another key parameter is the angle $\theta_B$ between the shock front normal and the external field. For example, non-relativistic perpendicular shocks ($\theta_B=\pi/2$) exhibit interesting non-stationary features that are discussed in section \ref{sec:shocksatb}. Also ultra-relativistic perpendicular shocks are not good Fermi accelerators, while parallel shocks ($\theta_B=0$) are \citep{Lemoine2006}. Finally, cases are when the energy of radiation is large enough to significantly enter the energy budget and alter the jump conditions. Such ``radiative'' shocks, collisionless or not, are encountered in astrophysical conditions and have equally been studied in literature \citep{Zeldovich,Bouquet2000}.

\subsection{Dominant instability and turbulence}\label{sec:weibel}
We here comment on the dominant instability at work in the shock formation, and on the resulting turbulence. Note that we are only interested in these processes to the extent that they participate in building up a shock. We thus won't give much detail on them and refer the reader to external references.

Counter-streaming instabilities are the reason why the overlapping region does not keep expanding, but turns turbulent instead \citep{Buneman1964}. Counter-streaming systems can be destabilized by many kind of instabilities, and the dominant one depends on the parameters of the problem (see \cite{BretPoPReview} for a review). For counter-streaming pair plasmas of nearly equal densities, the fastest growing modes are likely to be found with a wave vector normal to the flow \citep{BretPRL2008}. These modes are now referred to as filamentation, or Weibel, modes \citep{Weibel,Fried1959}.

In the case of electron/ion plasmas, where the relevant counter-streaming system consists in two opposite protons beams over a bath of hot electron, Weibel modes have also been found to govern the unstable spectrum  \citep{Shaisultanov2012}. ``Weibel shocks'' can therefore be found for a wide range of plasma types. Yet, it is well known that this instability is vulnerable to temperature \citep{Silva2002} or magnetization \citep{Godfrey1975,BretPoPMagne}. Simulations have been conducted where a flow-aligned magnetic field had been setup precisely to stabilize the Weibel modes \citep{SironiApj}. In this case, other unstable modes than Weibel take the lead. A turbulence is still generated which results in the formation of a shock.

Once the shock has been formed, the turbulence behind the front is maintained by the incoming upstream flow. This is how the collisionless shock eventually results in a self-sustained structure. Plasma instabilities keep arising near the shock front as a result of counter-streaming flows \citep{Karimabadi1991,Kato2010,Nakar2011,Lemoine2014EPL}.
But far downstream, where the upstream flow cannot arrive, the turbulence decays by phase mixing \citep{Chang2008,Lemoine2014}.

\section{Basic requirements for an EOS}\label{appA}
We here prove that it only takes an isotropic distribution function to have an equation of state like (\ref{eq:eos}). To this extent, we start performing a 1D analysis of the pressure in a gas with only 1 species. The calculation follows the derivation of \cite{feynman} (see \S39-2) who start from the definition of pressure.

Suppose a 1D gas along an $x$ axis, and an obstacle on the axis at $x=0$. Particles from its left bounce back against it, pushing it. The gas has the distribution function $f(x,v_x)$ (time dependence can be included). When bouncing, a particle with velocity $v_x$ transmits the momentum $dp=2mv_x$ to the obstacle. How many of them do so in a time $dt$? All the particles at distance $dl$ from the wall such that $dl<v_xdt$. The total amount of momentum transferred to the wall is thus,
\begin{equation}
\Delta p = \int_0^\infty dv_x ~ \int_{-v_x dt}^0 dx ~ f(x,v_x) ~ 2mv_x,
\end{equation}
where velocities are integrated from $0$ to $\infty$ only, because leftward particles do not hit the wall. We assume here that particles hitting the wall are by no means influenced by the others, which comes down to the perfect gas hypothesis.

For $dt\rightarrow 0$, $-v_x dt\rightarrow 0^-$, and
\begin{equation}
\int_{-v_x dt}^0 dx ~ f(x,v_x) ~ 2mv_x \sim v_x dt ~ f(0^-,v_x) ~ 2mv_x,
\end{equation}
so that,
\begin{equation}
\Delta p = \int_0^\infty dv_x ~ v_x dt ~ f(0^-,v_x) ~ 2mv_x \Rightarrow \frac{\Delta p}{dt} = 2 \int_0^\infty dv_x ~ v_x f(0^-,v_x) ~ mv_x.
\end{equation}
If $f(v_x)$ is isotropic, it is also an even function. Therefore, $2\int_0^{+\infty}dv_x = \int_{-\infty}^{+\infty}dv_x$ and
\begin{equation}
\frac{\Delta p}{dt} \equiv ~ \mathrm{Pressure} = \underbrace{ \int_{-\infty}^\infty dv_x ~ f(0^-,v_x) ~ v_x mv_x }_{\mathrm{Momentum~flux~at~}x=0^-}.
\end{equation}
As is well known, the pressure is thus related microscopically to the momentum flux at the left of the obstacle \citep{landau6}. The 3D result is straightforwardly derived, and reads (the wall is now $\perp$ to $x$),
\begin{equation}
P = \int d^3v ~ f(0^-,\mathbf{v}) ~ mv_x^2.
\end{equation}
If $f(\mathbf{v})$ is isotropic, then
\begin{equation}
\int d^3vf v_x^2=\int d^3vf v_y^2=\int d^3vf v_z^2~\Rightarrow ~ \int d^3vf v_x^2 = \frac{1}{3}\int d^3vf v^2.
\end{equation}
Hence,
\begin{equation}\label{eq:iso}
P = \frac{1}{3}\int d^3v ~ f(0^-,\mathbf{v}) ~ mv^2
= \frac{2}{3} \underbrace{ \int d^3v ~ f(0^-,\mathbf{v}) ~ \frac{1}{2}mv^2}_{\mathrm{Energy~density~}U}
 = \frac{2}{3} U,
\end{equation}
where the perfect gas hypothesis again is implied, if the energy density is to be solely made of kinetic energy. An equation of state of the form
\begin{equation}
 P=\frac{2}{3} U \equiv (\gamma-1)U,
\end{equation}
 is therefore recovered, where $\gamma$ is the polytropic index. It is clear that the ``2'' at the numerator comes from the necessity to cancel the $\frac{1}{2}$ factor in the kinetic energy formula, and that the ``3'' at the denominator is related to the number of dimensions of the system.

Therefore, $\gamma$ is still defined through
\begin{equation}
\gamma = 1+\frac{2}{d},
\end{equation}
where $d$ is the number of degrees of freedom of the particles. Here, we had $d=3$.

The same analysis can be conducted for the relativistic case, yielding the relativistic version of Eq. (\ref{eq:iso}),
\begin{equation}
P = \frac{1}{3}\int d^3v ~ f(0^-,\mathbf{v}) ~ \gamma(v)mv^2.
\end{equation}
In the ultra-relativistic limit, we can set $mv^2\sim mc^2$ in the integrand and obtain the ultra-relativistic equation of state \citep{landau1},
\begin{equation}
P = \frac{1}{3}U.
\end{equation}

%\bibliographystyle{jpp}
%\bibliography{BibBret}

\end{document}